\documentstyle[10pt]{article}
\newcommand{\bfp}{\mbox{\boldmath $p$}}

\newcommand{\bfk}{\mbox{\boldmath $k$}}

\def\nostrocostruttino#1\over#2{\mathrel{\mathop{\kern 0pt \rlap
{\hbox{$#1$}}} \hbox{\kern-.135em $#2$}}}

\newcommand{\NP}[1]{{\it Nucl.\ Phys.}\ {\bf #1}}
\newcommand{\ZP}[1]{{\it Z.\ Phys.}\ {\bf #1}}
\newcommand{\PL}[1]{{\it Phys.\ Lett.}\ {\bf #1}}
\newcommand{\PR}[1]{{\it Phys.\ Rev.}\ {\bf #1}}

\newcommand{\beq}{\begin{equation}}
\newcommand{\eeq}{\end{equation}}
\newcommand{\barr}{\begin{eqnarray}}
\newcommand{\earr}{\end{eqnarray}}
\newcommand{\ba}{\begin{array}}
\newcommand{\ea}{\end{array}}

\newcommand{\qq}{q\bar q}

\newcommand{\la}{\lambda}

\begin{document}

\begin{flushright}
DFTT 38/98 \\
INFNCA-TH9808 \\
hep-ph/9807366 \\
\end{flushright}

\vskip 0.3cm

\renewcommand{\thefootnote}{\fnsymbol{footnote}}

\begin{center}
{\bf Single spin asymmetries and vector meson production in DIS
\footnote{Talk delivered by M. Anselmino at the Workshop on ``Physics
and Instrumentation with 6-12 GeV Beams'', June 15-18, 1998, Jefferson Lab,
Newport News, Virginia.} \\ }

\vspace{0.4cm}
{\sf M. Anselmino$^1$ and F. Murgia$^2$} \\
\vspace*{0.4cm}
{$^1$Dipartimento di Fisica Teorica, Universit\`a di Torino and \\
      INFN, Sezione di Torino, Via P. Giuria 1, 10125 Torino, Italy \\
%\vskip 0.5cm
$^2$Dipartimento di Fisica, Universit\`a di Cagliari and \\
INFN, Sezione di Cagliari, CP 170, I-09042 Monserrato (CA), Italy} \\

\vspace*{0.5cm}
\end{center}

\begin{abstract}
We discuss possible measurements and origins of single spin asymmetries in 
DIS and of some unusual spin properties of vector mesons produced in $\ell N$, 
$\gamma N$ and $\gamma \gamma$ interactions. Such effects have already been 
observed in other processes. 
\end{abstract}

\vskip 12pt
\noindent
{\bf Single spin asymmetries in DIS}
\vskip 6pt
Single spin asymmetries in large $p_{_T}$ inclusive hadronic reactions are 
forbidden in leading-twist perturbative QCD, reflecting the fact that single 
spin asymmetries are zero at the partonic level and that collinear parton 
configurations inside hadrons do not allow single spin dependences. However, 
experiments tell us in several cases, \cite{ada1,ada2} that single spin 
asymmetries are large and indeed non negligible.

The usual arguments to explain this apparent disagreement between pQCD 
and experiment invoke the moderate $p_{_T}$ values of the data -- a few 
GeV, not quite yet in the true perturbative regime -- and the importance 
of higher-twist effects. Several phenomenological models have recently
attempted to explain the large single spin asymmetries observed in
$p^\uparrow p \to \pi X$ as twist-3 effects which might be due to intrinsic 
partonic $\bfk_\perp$ in the fragmentation \cite{col1} and/or distribution 
functions \cite{siv1}-\cite{noi}. 

A measurement of such single spin asymmetries also in Deep Inelastic 
Scattering (DIS) would add valuable information; a detailed analysis of 
single spin asymmetries in the inclusive, $\ell N^\uparrow \to \ell + jets$ 
and $\ell N^\uparrow \to hX$, reactions looking at possible origins and
devising strategies to isolate and discriminate among them can be found
in Ref. \cite{alm}.

We recall here only the main ideas and refer to Ref. \cite{alm} for
details and a full explanation of notations and the formalism. 

\vskip 6pt
\noindent
$a) \> \ell N^\uparrow \to \ell + 2\,jets$
\vskip 4pt
\nobreak

This case is the usual DIS on a nucleon $N$ with spin $S$ and one avoids 
any fragmentation effect by looking at the fully inclusive cross-section 
for the process $\ell N^\uparrow \to \ell + 2\, jets$, the 2 jets being the 
target and current ones. Within the QCD factorization theorem one has
\beq
\frac{d^2\sigma^{\ell + N,S \to \ell + X}} {dx \, dQ^2} = \sum_q
\int d^2\bfk_\perp \> \tilde f_{q/N}^{N,S}(x, \bfk_\perp) \>
\frac{d\hat\sigma^{q,P_q}}{d\hat t}(x, \bfk_\perp) \,. 
\label{gena}
\eeq

In this case the elementary interaction is a pure QED, helicity 
conserving one, $\ell q \to \ell q$, and $d\hat\sigma^{q,P_q}/d\hat t$
cannot depend on the quark polarization. Some spin dependence might only
remain in the distribution function $\tilde{f}_{q/N}^{N,S}$, due to intrinsic 
$\bfk_\perp$ effects \cite{siv1}-\cite{noi}, but is expected to be strongly 
suppressed by (necessary) initial state interactions. A dependence on 
the transverse nucleon spin of Eq. (\ref{gena}) is in principle possible, 
but would be very surprising and intriguing \cite{dra}.

\vskip 6pt
\noindent
$b) \> \ell N^\uparrow \to h + X \> (2\,jets, \> \bfk_\perp \not= 0)$
\vskip 4pt

One looks for a hadron $h$, with transverse momentum $\bfk_\perp$, inside 
the quark current jet; the final lepton may or may not be observed.
The elementary subprocess is $\ell q \to \ell q$ and one has \cite{col2,col3}
\barr
& & \frac{E_h \, d^5\sigma^{\ell + N^\uparrow \to h + X}} 
{d^{3} \bfp_h d^2 \bfk_\perp} 
- \frac{E_h \, d^5\sigma^{\ell + N^\downarrow \to h + X}} 
{d^{3} \bfp_h d^2 \bfk_\perp} \label{coll} \\
&=& \sum_q \int \frac {dx} {\pi z} \>    
\Delta_{_T} q(x) \> \Delta_{_N} \hat\sigma^q (x, \bfk_\perp) \,
\left[ \tilde D^h_{q^\uparrow}(z, \bfk_\perp)
- \tilde D^h_{q^\uparrow}(z, - \bfk_\perp) \right]
\nonumber
\earr
where $\Delta_{_T}q$ (or $h_1$) is the polarized number density for 
transversely spinning quarks $q$ and $\Delta_{_N} \hat\sigma^q$ is the 
elementary cross-section double spin asymmetry
\beq
\Delta_{_N} \hat\sigma^q = {d\hat \sigma^{\ell q^\uparrow \to 
\ell q^\uparrow} \over d\hat t} - {d\hat \sigma^{\ell q^\uparrow \to 
\ell q^\downarrow} \over d\hat t} \,\cdot
\label{del}
\eeq

In Eq. (\ref{coll}) we have neglected the $\bfk_\perp$ effect in the 
distribution function, which can be done once the asymmetry discussed
in $a)$ turns out to be negligible. We are then testing directly the 
mechanism suggested in Ref. \cite{col1} and a non zero value of
the l.h.s. of Eq. (\ref{coll}) would be a decisive test in its favour 
and would allow an estimate of the new function $[\tilde D^h_{q^\uparrow}
(z, \bfk_\perp) - \tilde D^h_{q^\uparrow}(z, - \bfk_\perp)$]. 
Notice that even upon integration over
$d^2\bfk_\perp$ the spin asymmetry of Eq. (\ref{coll}) might survive,
at higher twist order $k_\perp/p_{_T}$, due to some $\bfk_\perp$ dependence 
in $\Delta_{_N} \hat\sigma^q$.

Several other cases are considered in Ref. \cite{alm}.

\vskip 12pt
\noindent
{\mbox{\boldmath $\rho^{\,}_{1,-1}(V)$}} {\bf in}
{\mbox{\boldmath $\ell N \to V + X, \> \gamma N \to V + X$}} {\bf and}
{\mbox{\boldmath $\gamma \gamma \to V + X$}}
\vskip 6pt
\noindent
In Ref. \cite{akp} it was suggested how the coherent fragmentation of
$\qq$  pairs created in $e^+ e^- \to \qq \to V + X$ processes might lead
to non zero  values of the off-diagonal element $\rho_{1,-1}$ of the
helicity density  matrix of the vector mesons $V$; in Ref. \cite{abmq}
actual predictions were  given for several spin 1 particles produced at
LEP energies in two jet events,  provided they carry a large fraction
$x_{_E}$ of the parent quark energy and have  a small intrinsic $\bfk_\perp$,
{\it i.e.} they are collinear with the parent jet.

The values of $\rho_{1,-1}(V)$ are related to the values of the off-diagonal
helicity density matrix element $\rho_{+-;-+}(\qq)$ of the $\qq$ pair,
generated in the $e^- e^+ \to \qq$ process \cite{abmq}:
\beq
\rho^{\,}_{1,-1}(V) \simeq [1 - \rho^{\,}_{0,0}(V)] \>
\rho_{+-;-+}(\qq) \label{old}
\eeq
where the value of the diagonal element $\rho^{\,}_{0,0}(V)$ can be taken from
data. The values of $\rho_{+-;-+}(\qq)$ depend on the elementary short distance
dynamics and can be computed in the Standard Model. Thus, a measurement
of $\rho_{1,-1}(V)$, is a further test of the constituent dynamics, more
significant than the usual measurement of cross-sections in that it depends
on the product of different elementary amplitudes, rather than on squared
moduli:
\beq
\rho^{\,}_{+-;-+}(\qq) = {1\over 4N_{\qq}} \sum_{\la^{\,}_{-}, \la^{\,}_{+}}
M^{\,}_{+-;\la^{\,}_{-} \la^{\,}_{+}} \> M^*_{-+; \la^{\,}_{-} \la^{\,}_{+}}
\,, \label{rhoz}
\eeq
where the $M$'s are the helicity amplitudes for the $e^- e^+ \to \qq$ process
and $N_{\qq}$ is the normalization factor. With unpolarized $e^+$ and 
$e^-$, at LEP energy, $\sqrt s = M_{_Z}$, one has \cite{abmq}
\beq
\rho^{\,}_{+-;-+}(\qq) \simeq
\rho^{Z}_{+-;-+}(\qq) \simeq {1\over 2} \> {(g^2_{_V} - g^2_{_A})_q \over
(g^2_{_V} + g^2_{_A})_q} \, {\sin^2\theta \over 1+ \cos^2\theta} \, \cdot
\label{rhozap}
\eeq

Eq. (\ref{old}) is in good agreement with OPAL Collaboration data on
$\phi$, $D^*$ and $K^*$, including the $\theta$ dependence induced by Eq.
(\ref{rhozap}) \cite{opal1, opal2}; however, no sizeable value of
$\rho_{1,-1}(V)$ for $V= \rho, \phi$ and $K^*$ was observed by DELPHI
Collaboration \cite{delphi}. Further tests are then necessary.

We consider here other interactions -- of interest for the Jefferson Lab 
program -- in which the value of $\rho_{1,-1}(V)$ could be measured, 
namely $\gamma N \to VX$, $\ell N \to \ell VX$ and possibly $\gamma \gamma 
\to VX$,  with $V= \phi, D^*$ or $B^*$. 
The choice of a heavy vector meson implies the dominance in each of these 
cases of particular elementary hard contributions, $\gamma g \to \qq$, 
$\gamma^* g \to \qq$ and $\gamma \gamma \to \qq$, with $q = s, c $ or $b$. 

The hadronization process -- the fragmentation of a $\qq$
pair -- is then similar to the one occurring in $e^+e^-$ annihilations;
however, the value of $\rho_{1,-1}(V)$ in these cases should be different
from that observed in $e^-e^+ \to VX$ at LEP, due to a different underlying
elementary dynamics, {\it i.e.} a different value of $\rho^{\,}_{+-;-+}(\qq)$.
A measurement of $\rho_{1,-1}(V)$ in agreement with our predictions in these
other processes would be an unambiguous test of both the quark hadronization
mechanism and the real nature of the constituent interactions.

Details of the calculation can be found in Ref. \cite{abmp}; we find
again Eq. (\ref{old}) with, neglecting quark masses and for real photons:
\beq
\rho_{+-;-+}(\qq) = \rho^{\gamma, g}_{+-;-+}(\qq;\theta^*) = {1 \over 2} \>
{\sin^2\theta^* \over 1 + \cos^2 \theta^*} \,\cdot
\label{rhopg}
\eeq
This value of $\rho^{\gamma, g}_{+-;-+}(\qq)$ is the same for all the 
possible elementary processes initiated by a real photon or a real gluon,
{\it i.e.} also for resolved photon contributions; $\theta^*$ is the 
production angle of $q$ and $V$ in the partonic {\it c.m.} frame. 

The situation is different and potentially very interesting in case 
of DIS, because the value of $\rho^{\gamma^* g \to \qq}_{+-;-+}$ strongly
depends on the DIS variables \cite{abmp}
\beq
\rho^{\gamma^* g \to \qq}_{+-;-+}(\qq;y,z,\theta^*) =
{\sin^2 \theta^* \over 2(1 + \cos^2 \theta^*)} \> 
{1 - A(y,z) \over 1 + A(y,z) \sin^2\theta^*/(1+\cos^2\theta^*)}
\label{ayz}
\eeq
where
\beq
A(y,z) = {8z(1-z)(1-y) \over [(1 - z)^2 + z^2] \> [1 + (1-y)^2]} \,\cdot
\label{defa}
\eeq

Insertion of Eqs. (\ref{ayz}) and (\ref{defa}) into Eq. (\ref{old})
gives our prediction for $\rho_{1,-1}(V)$ in DIS, and its dependence on the 
variables $y = Q^2/(sx)$, $z = x/x_g$, the production angle $\theta^*$ of $V$ 
in the $\gamma^* g$ $c.m.$ frame and the measured value of $\rho_{0,0}(V)$.

The two issues considered here are only two examples of interesting spin 
physics possible at Jefferson Lab; many other spin observables and spin
effects should be measurable and should allow to collect precious 
information on subtle and little known properties both of quark distributions
and fragmentation.  

\vskip 6pt
{\bf Acknowledgements}
\vskip 4pt
\noindent
One of us (M.A.) would like to thank the organizers of the Workshop for the 
kind invitation and Jefferson Lab for financial support.

\end{document}